\documentclass[prb]{revtex4}

\usepackage{graphicx}
\usepackage{dcolumn}
\usepackage{bm}

\begin{document}

\preprint{APS/123-QED}

\title{Hamiltonian formalism of the Landau-Lifschitz equation for a spin chain with full anisotropy}

\author{Nian-Ning Huang}
\author{Hao Cai}%
 \email{caihao2001@hotmail.com}
\author{Tian Yan}
\author{Fan-Rong Xu}
\affiliation{%
Department of Physics, Wuhan University, Wuhan 430072, P.R.China
}%

\date{\today}

\begin{abstract}
The Hamiltonian formalism of the Landau-Lifschitz equation for a
spin chain with full anisotropy is formulated completely, which
constructs a stable base for further investigations.


\end{abstract}

\pacs{Valid PACS appear here}
\maketitle

\section{Introduction}

The Landau-Lifschitz(L-L) equation for a spin chain with
full anisotropy is one of the most important completely integrable
nonlinear evolution equations\cite{1}. The compatibility pair were
first given by E.K.Sklyanin\cite{2} in terms of elliptic
functions. The Hamiltonian formalism was also examined, but the
reduction procedure and results have some open points so that the
final results are not conclusive\cite{2}. On the other hand, there
were some works trying to solve the equation based upon these
compatibility pair, but exact solutions were not found until
now\cite{3,4}. Because these problems are more basic, it is worth
to putting a lot of work. In this paper the Hamiltonian formalism
is formulated to provide a base for further investigations.

\section{Landau-Lifschitz equation}

The L-L equation for anisotropic spin chain is in the form of
\begin{equation}\label{1}
    S_t=S\times S_{xx}+S\times JS,~~|S|=1
\end{equation}
in which $J={\rm diag}(J_1,J_2,J_3)$, $J_1\leq J_2\leq J_3$,
describing nonlinear spin waves propagating in a
direction orthogonal to the anisotropic axis, and the suffices $x$ and
$t$ denote the corresponding partial derivatives. Though spin is a
quantum quantity satisfying quantum bracket relation, but if one
introduces Lie-Poisson bracket for classical spin, all the procedure of
disposal is the same as the usual classical one\cite{1}. The
Lie-Poisson bracket is given by
\begin{equation}\label{2}
    \{S_a(x),S_b(y)\}=-\epsilon_{abc}S_c(x)\delta(x-y)
\end{equation}
where $\epsilon_{abc}$ is a fully anti-symmetric tensor, $a,b,c =1,2,3$.
By using it we write eq.(\ref{1}) in the form of the canonical
equation
\begin{equation}\label{3}
    \partial_tS_a(x)=\{S_a(x),H\}
\end{equation}
in which the Hamiltonian is
\begin{equation}\label{4}
    H=\int_{-\infty}^{\infty}dx{\cal H}(x),~~
    {\cal H}(x)=\frac{1}{2}\sum_a\left[\big(\partial_xS_a(x)\big)^2-J_aS_a(x)^2\right]
\end{equation}

\section{Jost solutions}

The compatibility pair of the equation are a couple of $2\times 2$
matrices given by E.K.Sklyanin\cite{2},
\begin{equation}\label{5}
    L=-i\sum_a\omega_a(\lambda)S_a\sigma_a
\end{equation}
and
\begin{equation}\label{6}
    M=-i\sum_{a,b,c}\omega_a(\lambda)S_bS_{c x}\sigma_a \epsilon_{abc}
+i\sum_{a,b,c}\omega_b(\lambda)\omega_c(\lambda)S_a\sigma_a
|\epsilon_{abc}|
\end{equation}
in which
\begin{equation}\label{7}
    \omega_1(\lambda)=\rho \frac{1}{{\rm sn} (\lambda,\kappa)}, \ \
\omega_2(\lambda)=\rho \frac{{\rm dn}(\lambda,\kappa)}{{\rm sn}
(\lambda,\kappa)}, \ \ \omega_3(\lambda)=\rho \frac{{\rm
cn}(\lambda,\kappa)}{{\rm sn} (\lambda,\kappa)}
\end{equation}
and ${\rm sn}(\lambda,\kappa)$, etc., are elliptic functions with
modulus $\kappa$:
\begin{equation}\label{8}
    \kappa=\sqrt{\frac{J_2-J_1}{J_3-J_1}},~~
\rho=\frac{1}{2}\sqrt{J_3-J_1}
\end{equation}
Since the coefficients $\omega_a(\lambda)$ are double-periodic
functions of the parameter $\lambda$:
\begin{equation}\label{9}
    \omega_a(\lambda+4mK+i4nK')=\omega_a(\lambda)
\end{equation}
where $n,m$ are integers, $K$ and $K'$ are quarter-periods. It
is sufficient to consider $\lambda$ in the fundamental period
parallelogram $\Omega$:
\begin{equation}\label{10}
    \Omega:~|{\rm Re}\lambda|< 2K,~~|{\rm Im}\lambda|< 2K'
\end{equation}

The first compatibility equation is now written as
\begin{equation}\label{11}
    \partial_x F(x,\lambda)=L(x,\lambda)F(x,\lambda)
\end{equation}
In the limit of $|x|\to \pm\infty$, $S\to (0,0,1)$, and the
asymptotic solution corresponding is
$E(x,\lambda)=e^{-i\omega_3x\sigma_3}$. The Jost solutions of
eq.(11) are then defined by
\begin{eqnarray}
\Psi(x,\lambda)=\big(\tilde{\psi}(x,\lambda),\psi(x,\lambda)\big)\to e^{-i\omega_3x\sigma_3},&~&
\mbox{as}~x\to\infty\nonumber\\
\Phi(x,\lambda)=\big(\phi(x,\lambda),\tilde{\phi}(x,\lambda)\big)\to e^{-i\omega_3x\sigma_3},&~&
\mbox{as}~x\to-\infty\label{12}
\end{eqnarray}
The monodromy matrix $T(\lambda)$ is given by
\begin{equation}\label{13}
    \Phi(x,\lambda)=\Psi(x,\lambda)T(\lambda),~~
T(\lambda)=\left(\begin{array}{rr}a(\lambda)&-\tilde{b}(\lambda)\\
[4pt] b(\lambda)&\tilde{a}(\lambda)\end{array}\right)
\end{equation}
Furthermore, from the periodic properties of ${\rm sn}(\lambda)$ and
${\rm sn}(\bar{\lambda})=\overline{{\rm sn}(\lambda)}$, etc.,
the compatibility pair (5) and (6) have the following reduction
transformation properties:
\begin{equation}\label{14}
    L(\lambda+2K)=\sigma_3L(\lambda)\sigma_3,~~
M(\lambda+2K)=\sigma_3M(\lambda)\sigma_3
\end{equation}
\begin{equation}\label{15}
    \overline{L(\bar{\lambda}+i2K')}=\sigma_3L(\lambda)\sigma_3,~~
\overline{M(\bar{\lambda}+i2K')}=\sigma_3M(\lambda)\sigma_3
\end{equation}

\section{Lie-Poisson bracket}

From the first compatibility equation, it is found that
\begin{equation}\label{16}
    \frac{\delta T(\lambda)}{\delta S_a(z)}=
{\Psi}^{-1}(z,\lambda)\left(i\omega_a\sigma_a\right)
{\Phi}(z,\lambda)
\end{equation}
and
\begin{equation}\label{17}
    \frac{\delta T^{-1}(\lambda)}{\delta S_b(z)}=
-{\Phi}^{-1}(z,\lambda)\left(i\omega_b\sigma_b\right)
{\Psi}(z,\lambda)
\end{equation}
Defining$^{5}$
\begin{equation}\label{18}
    \{T(\lambda){\stackrel{\displaystyle{\otimes}}{,}}
T^{-1}(\lambda')\}_{ik,jl}=\{T(\lambda)_{ij},T^{-1}(\lambda')_{kl}\}
\end{equation}
the Lie-Poisson bracket of the monodromy matrix is now simply
given in the form
\begin{equation}\label{19}
    \{T(\lambda){\stackrel{\displaystyle{\otimes}}{,}}
T^{-1}(\lambda')\}=-\epsilon_{abc}\int dx\frac{\delta
T(\lambda)}{\delta S_a(x)}\otimes \frac{\delta
T^{-1}(\lambda')}{\delta S_b(x)} S_c(x)
\end{equation}
in which the symbol $\otimes$ in the right hand side is the usual direct product.
After substituting eqs.(\ref{16}) and (\ref{17}), the explicit expression of
eq.(19) is
\begin{equation}\label{20}
    \{T(\lambda){\stackrel{\displaystyle{\otimes}}{,}} T^{-1}(\lambda')\}=\int dx
\Psi^{-1}(x,\lambda)\Phi^{-1}(x,\lambda')R\Phi(x,\lambda)\Psi(x,\lambda')
\end{equation}
where
\begin{eqnarray}
    R&=&S_3\{i\omega_1\omega'_2\sigma_1\otimes
i\sigma_2-i\omega_2\omega'_1i\sigma_2\otimes \sigma_1\}
+S_1\{i\omega_2\omega'_3i\sigma_2\otimes
\sigma_3-i\omega_3\omega'_2\sigma_3\otimes i\sigma_2\}\nonumber\\
&~&~~+S_2\{-\omega_3\omega'_1\sigma_3\otimes
\sigma_1+\omega_1\omega'_3\sigma_1\otimes \sigma_3\}\label{21}
\end{eqnarray}

The eq.(\ref{20}) has a simple result if the integrand in the right hand
side is a full derivative of some function with respect to $x$.
In order to do so,
we take into account of
\begin{equation}\label{23}
    \partial_x\Big(\Psi^{-1}(x,\lambda)\sigma_{\alpha}\Psi(x,\lambda')\otimes'
\Phi^{-1}(x,\lambda')\sigma_{\alpha}\Phi(x,\lambda)\Big)
\end{equation}
where $\sigma_0=I$, and $\sigma_a$ is Pouli's matrix for
$a=1,2,3$. Here another type of direct product is introduced
\begin{equation}\label{24}
    A_{im}B_{lj}=(A\otimes' B)_{il,jm}
\end{equation}
Using eqs.(\ref{5}) and (\ref{11}), it is obvious that
\begin{equation}\label{25}
    \Psi^{-1}(x,\lambda)\Phi^{-1}(x,\lambda')W_{\alpha}\Phi(x,\lambda)\Psi(x,\lambda')
\end{equation}
in which
\begin{equation}\label{26}
    W_0=i( \omega_a- \omega'_a)S_a(\sigma_a\otimes'
I-I\otimes'\sigma_a)
\end{equation}
\begin{equation}\label{27}
    W_a=iS_b( \omega_b\sigma_b\sigma_a+
\omega'_b\sigma_a\sigma_b)\otimes'\sigma_a +iS_b\sigma_a\otimes'(
\omega'_b\sigma_b\sigma_a+ \omega_b\sigma_a\sigma_b)
\end{equation}
In fact, eq.(\ref{20}) can be expressed as a linear combination of terms
in eq.(\ref{23}), since $R$ in eq.(\ref{21}) is expressed as a linear
combination of $W_{\alpha}$ in eqs.(\ref{26}) and (\ref{27}):
\begin{equation}\label{28}
    R=f_0W_0+f_3W_3+f_1W_1+f_2W_2
\end{equation}
Writing $R$ and $W_{\alpha}$ in bigger matrices, e.g.
$4\times4-$matrices, and comparing the corresponding matrix
elements, a group of equations for $f_{\alpha}$ are given in the Appendix A.

Eq.(\ref{20}) is then
\begin{equation}\label{29}
    \{T(\lambda){\stackrel{\displaystyle{\otimes}}{,}}
T^{-1}(\lambda')\}=\sum_{\alpha}f_{\alpha}\Delta_{\alpha}
\end{equation}
where
\begin{equation}\label{30}
    \Delta_{\alpha}\equiv\Psi^{-1}(x,\lambda)\sigma_{\alpha}\Psi(x,\lambda')\otimes'
\Phi^{-1}(x,\lambda')\sigma_{\alpha}\Phi(x,\lambda)\Big|_{x=-L}^{x=L}
\end{equation}

\section{Explicit expression of Lie-Poisson bracket}

On the other hand, we have
\begin{equation}\label{33}
    \sum_{\alpha=0}^3
f_{\alpha}\Delta_{\alpha}=f_0(\Delta_0+\Delta_3+\Delta_1+\Delta_2)+(f_3-f_0)\Delta_3
+(f_1-f_0)\Delta_1+(f_2-f_0)\Delta_2
\end{equation}
Since $\{T(\lambda){\stackrel{\displaystyle{\otimes}}{,}}T^{-1}(\lambda')\}$ is definite
and only the differences $f_3-f_0$, $f_1-f_0$, $f_2-f_0$ can be determined from eqs.(\ref{a4})$\sim$(\ref{a6}),
we should see that $\Delta_0+\Delta_3+\Delta_1+\Delta_2=0$, which means that the
value of $f_0$ is of no importance and may be assumed to be $0$. Thus we have
\begin{equation}\label{34}
    \{T(\lambda){\stackrel{\displaystyle{\otimes}}{,}} T^{-1}(\lambda')\}=
f_3(\Delta(b)-\Delta(b0))+f_1(-\Delta(b)+\Delta(b1))+
f_2(-\Delta(b)-\Delta(b1))
\end{equation}
in which
\begin{equation}\label{35}
    f_3=\frac{\omega_1\omega'_2+\omega_2\omega'_1}{2(\omega_3-\omega_3')},~~
f_1=\frac{\omega_2\omega'_3+\omega_3\omega'_2}{2(\omega_1-\omega_1')},~~
f_2=\frac{\omega_3\omega'_1+\omega_1\omega'_3}{2(\omega_2-\omega'_2)}
\end{equation}
and the explicit expressions of $\Delta(b)$, $\Delta(b0)$ and $\Delta(b1)$ are given in the Appendix B.

Because of properties shown in (\ref{14}), $\lambda$ can be restricted in the region
$\Omega_{+}$:
\begin{equation}\label{36}
    \Omega_{+}: \ 0<{\rm Re}\lambda<2K,~~0<{\rm Im}\lambda<2K'
\end{equation}
In this restriction, $\Delta(b1)$ has no contribution, and (\ref{34})
reduces to
\begin{eqnarray}
&&    \{T(\lambda){\stackrel{\displaystyle{\otimes}}{,}} T^{-1}(\lambda')\}= f_3(\Delta(b)-\Delta(b0))\nonumber\\
&&~~=-\frac{\omega_1\omega'_2+\omega_2\omega'_1}{2}
    \left(\!\!\begin{array}{cccc}
0&-\frac{1}{\omega_3-\omega'_3+i0}a\tilde{b}'&
-\frac{1}{\omega_3-\omega'_3+i0}\tilde{a}'\tilde{b}&0\\
-\frac{1}{\omega_3-\omega'_3+i0}b'a&0&
i2\pi\delta(\omega_3-\omega'_3)|a|^2&\frac{1}{\omega_3-\omega'_3-i0}\tilde{b}a'\\
-\frac{1}{\omega_3-\omega'_3+i0}b\tilde{a}'&-i2\pi\delta(\omega_3-\omega'_3)|a|^2&0&
\frac{1}{\omega_3-\omega'_3-i0}\tilde{b}'\tilde{a}\\
0&\frac{1}{\omega_3-\omega'_3-i0}a'b&\frac{1}{\omega_3-\omega'_3-i0}\tilde{a}b'&
0
\end{array}\!\!\right)\label{37}
\end{eqnarray}
where $\omega_j,~\omega'_j$ mean $\omega_j(\lambda),~\omega_j(\lambda')$.

From eq.(\ref{37}), there are
\begin{eqnarray}
    \{a(\lambda),\,b(\lambda')\}
    &=&\frac{\omega_1\omega'_2+\omega_2\omega'_1}{2}\frac{1}{\omega_3-\omega_3'+i0}ab'\nonumber\\
    &=&\frac{\omega_1\omega'_2+\omega_2\omega'_1}{2(\omega-\omega')}ab'
-\frac{\omega_1\omega'_2+\omega_2\omega'_1}{2}i\pi\delta(\omega_3-\omega'_3)ab'\label{39}
\end{eqnarray}
\begin{eqnarray}
    \{\tilde{a}(\lambda),\,b(\lambda')\}
    &=&-\frac{\omega_1\omega'_2+\omega_2\omega'_1}{2}\frac{1}{\omega_3-\omega_3'-i0}\tilde{a}b'
    \nonumber\\
&=&-\frac{\omega_1\omega'_2+\omega_2\omega'_1}{2(\omega-\omega')}ab'
-\frac{\omega_1\omega'_2+\omega_2\omega'_1}{2}i\pi\delta(\omega_3-\omega'_3)ab'\label{40}
\end{eqnarray}
and then
\begin{equation}\label{41}
    \{|{a}(\lambda)|^2,\,b(\lambda')\}
    =-i2\pi\delta(\omega_3-\omega'_3)\omega_1\omega_2|a|^2b'
\end{equation}

Furthermore, in the restriction $\Omega_{+}$, there are
\begin{equation}\label{42}
    \delta(\omega_3-\omega_3')=\frac{1}{\frac{d\omega_3(\lambda)}{d\lambda}}\delta(\lambda-\lambda')
\end{equation}
as $\lambda=\lambda'$, and
\begin{equation}\label{43}
    \frac{\omega_1\omega'_2+\omega_2\omega'_1}{2}=\rho^2\frac{{\rm
dn}(\lambda)}{{\rm
sn}^2(\lambda)}=-\rho\frac{d\omega_3(\lambda)}{d\lambda}
\end{equation}
As a result, eqs.(\ref{39}), (\ref{40}) and (\ref{41}) become
\begin{equation}\label{44}
    \{a(\lambda),\,b(\lambda')\}=\frac{\omega_1\omega'_2+\omega_2\omega'_1}{2(\omega-\omega')}
a(\lambda)b(\lambda')
+i\pi\rho\delta(\lambda-\lambda')a(\lambda)b(\lambda')
\end{equation}
\begin{equation}\label{45}
    \{\tilde{a}(\lambda),\,b(\lambda')\}=-\frac{\omega_1\omega'_2+\omega_2\omega'_1}{2(\omega-\omega')}
\tilde{a}(\lambda)b(\lambda')
+i\pi\rho\delta(\lambda-\lambda')\tilde{a}(\lambda)b(\lambda')
\end{equation}
and
\begin{equation}\label{46}
    \{|{a}(\lambda)|^2,\,b(\lambda')\}=
i2\pi\rho\delta(\lambda-\lambda')|{a}(\lambda)|^2b(\lambda')
\end{equation}

\section{Action-angle variables in continuous spectrum}

As known in inverse scattering transform, $a(\lambda)$ and
$\tilde{a}(\lambda)$ are independent of $t$, and the phase of
$b(\lambda)$ and $\tilde{b}(\lambda)$ is a function of $t$, which
is determined by the asymptotic form of $M$ in eq.(\ref{6}). We have
\begin{equation}\label{47}
    b(t,\lambda)=b(0,\lambda)e^{-i4\omega_1\omega_2t}
\end{equation}
and then the angle variable is defined as
\begin{equation}\label{48}
    Q(\lambda)={\rm arg}\,b(\lambda)=\frac{1}{i}{\rm
ln}\,{b(\lambda)}
\end{equation}
The action variable $P(\lambda)$ is a function of $a(\lambda)$ and
$\tilde{a}(\lambda)$, and usually assumed to be
\begin{equation}\label{49}
    P(\lambda)=F(|a(\lambda)|^2)
\end{equation}
where $F$ is an unknown function. As these two variables are
canonical variables, it should be
\begin{equation}\label{50}
    \{P(\lambda),Q(\lambda')\}=-\delta(\lambda-\lambda')
\end{equation}
By eq.(42), we find
\begin{equation}\label{51}
    \{F(|a(\lambda)|^2),\,Q(\lambda')\}=
F'(|a(\lambda)|^2)2\pi\rho\delta(\lambda-\lambda')|a(\lambda)|^2
\end{equation}
where $F'$ is derivative of $F$ with respect of its argument.
Comparing it with eq.(46), we obtain
\begin{equation}\label{52}
    F'(|a(\lambda)|^2)2\pi\rho|a(\lambda)|^2=-1
\end{equation}
and thus
\begin{equation}\label{53}
    P(\lambda)=F(|a(\lambda)|^2)=-\frac{1}{2\pi\rho}{\rm
ln}\,{|a(\lambda)|^2}
\end{equation}
Eq.(44) yields
\begin{equation}\label{54}
    Q(\lambda,t)=Q(\lambda,0)-4\omega_1\omega_2
\end{equation}
Hence the Hamiltonian is
\begin{equation}\label{55}
    H=\int_{0}^{2K}d\lambda 4\omega_1\omega_2P(\lambda)=
-\frac{2}{\pi}\int_{0}^{2K}d\lambda\, \frac{\rho{\rm
dn}(\lambda)}{{\rm sn}^2(\lambda)}{\rm ln}\,{|a(\lambda)|^2}
\end{equation}

Therefore, the Hamiltonian has two kinds of expressions: one is an integral
with respect to $x$ in eq.(\ref{4}), and the other is an integral with
respect to the spectral parameter in eq.(\ref{55}). Now it is necessary
to derive a conservative quantity which has two integral forms
compatible with the Hamiltonian.

\section{Conservative quantities}

In the inverse scattering transform, the conservative quantities are
derived from the asymptotic form of the first compatibility equation.
In the limit $|\lambda|\to 0$ or $|k|\to\infty$ ($k=\rho\lambda^{-1}$),
by using the asymptotic expansion of
elliptic functions, we have
\begin{equation}\label{56}
    L\to-i(k+k^{-1}r_a)S_{a}\sigma_{a}+\cdots
\end{equation}
and
\begin{equation}\label{57}
    \{r_1,r_2,r_3\}=4\rho^2\frac{1}{6}\{(1+\kappa^2),(1-2\kappa^2),(-2+\kappa^2)\}
\end{equation}
From the first compatibility equation in this limit, writing
\begin{equation}\label{58}
    \tilde{\psi}_2(x,k)=e^{ik x+g}
\end{equation}
and expanding
\begin{equation}\label{59}
    g_x=\eta_0+(i2k)^{-1}\eta_1+(i2k)^{-2}\eta_2+...
\end{equation}
we obtain
\begin{equation}\label{60}
    \eta_0=S_{3x} +\frac{(-iS_1+S_2)_x}{-iS_1+S_2}(1-S_3)
\end{equation}
and
\begin{equation}\label{61}
    -2\eta_1=\eta_{0x}+2\eta_0^2
-\frac{(-iS_1+S_2)_x}{-iS_1+S_2}(\eta_0)+2r_aS_a^2
\end{equation}
etc.\cite{6}. In general, $\eta_0\neq 0$. This situation appears
also in the case of isotropic spin chain. In that case, an
additional phase in the transmission coefficient $a(k)$ was
introduced\cite{5} to cancel the non-vanishing $\eta_0$. However,
Takhtajan and Zakharov pointed out that this is unreasonable$^{7}$.
Any way, $\eta_1$ in eq.(\ref{61}) does not
give an expression compatible with the Hamiltonian in
eq.(\ref{4}).

It was formerly shown that the gauge
equivalence between the isotropic spin chain and the nonlinear
Schr\"odinger equation, which means, by choosing a suitable gauge, the
gauge-transformed compatibility pair of isotropic spin chain has
the same form of that of NLS equation. Therefore, the conservative
quantities for the isotropic spin chain can be derived from those
for the NLS equation by revised gauge transformation and all
results expected are naturally found\cite{7}.

After Takhtajan and Zakharov\cite{7}, it was tried to find some
equations that are gauge-equivalent to the L-L equations for a
spin chain with axial symmetry. However, such equations seem not
existent. From a careful analysis of the gauge equivalence between the
isotropic spin chain and the NLS equation,
only the leading terms of the first one of compatibility pair are essentials,
while other terms corresponding and the second one of compatibility pair are of no
importance. That is, a gauge is chosen such that it turns the spin
in the first order of spectral parameter of the first one of
compatibility pair into the $3-$ axis in the spin space\cite{6}.

The explicit expression of the gauge $B$ was given(see Ref.[6]).
After the gauge transformation, eq.(\ref{56}) turns to
\begin{equation}\label{62}
    L'=-i(k+2r_aS_a^2)\sigma_3+U+k^{-1}V+\cdots
\end{equation}
where $U\equiv B_xB^{-1}=\left(\begin{array}{cc}0&u\\
-\bar{u}&0\end{array}\right)$ and $V$ is an $2\times 2$ matrix with
vanishing diagonal elements.

Replacing $L$ by $L'$, the similar procedure for eq.(\ref{60}) and (\ref{61}) yields
\begin{equation}\label{63}
    \eta'_0=0,~~\eta'_1=-|u|^2+2(r_1S_1^2+r_2S_2^2+r_3S_3^2)
\end{equation}
Noticing (\ref{8}), there are
\begin{equation}\label{64}
    \{r_1,r_2,r_3\}=\frac{1}{6}\{J_3+J_2-2J_1,J_3-2J_2+J_1,
-2J_3+J_1+J_2\}\sim-\frac{1}{2}\{J_1,J_2,J_3\}
\end{equation}
since common constant is immaterial\cite{8}. As shown in the case of
isotropic spin chain$^{7}$
\begin{equation}\label{65}
    4|u|^2=-S_{a x}S_{a x}
\end{equation}
We finally obtain
\begin{equation}\label{66}
    \eta'_1=\frac{1}{2}\left(S_{ax}S_{ax}-(J_1S_1^2+J_2S_2^2+J_3S_3^2)\right)
\end{equation}
which is just the density of the Hamiltonian in eq.(4).

\section{Expression of $a(\lambda)$}

Eq.(\ref{63}) indicates $a(k)\to 0$ as $k\to\infty$, so that in this
domain
\begin{equation}\label{67}
    {\rm ln}a(k)=\frac{1}{i\pi}\int dk' \frac{{\rm
ln}|a(k')|^2}{k'-k}
\end{equation}
We have
\begin{equation}\label{68}
    {\rm ln}a(k)=I_0+I_1(i2k)^{-1}+\cdots
\end{equation}
\begin{equation}\label{69}
    I_0=0,~~I_1=-\frac{2}{\pi}\int dk' {{\rm ln}|a(k')|^2}
\end{equation}
In terms of $\lambda$($k=2\rho\lambda^{-1}$), eq.(\ref{67}) is re-written in the domain of
$\lambda\approx 0$
\begin{equation}\label{70}
    {\rm ln}a(\lambda)=\frac{1}{i\pi}\int d\lambda' \frac{{\rm
ln}|a(\lambda')|^2}{(\lambda'-\lambda)}\frac{\lambda}{\lambda'}
\end{equation}
so that
\begin{equation}\label{71}
    I_1=-\frac{2}{\pi}\int d\lambda'\frac{\rho}{\lambda'^2} {\rm
ln}|a(\lambda')|^2
\end{equation}
It is equal to eq.(\ref{55}) in this domain, since ${\rm
dn}(\lambda)\approx 1$ and ${\rm sn}(\lambda)\approx
\lambda^{-1}$.

Since the general formula in this case is invariant with
double-periodic transformation, eq.(\ref{70}) may be rewritten as
\begin{equation}\label{72}
    {\rm ln}a(\omega_3(\lambda))=\frac{1}{i\pi}\int_{\Omega_{+}} d
\omega_3(\lambda') \frac{{\rm
ln}|a(\omega_3(\lambda'))|^2}{\omega_3(\lambda')-\omega_3(\lambda)}
\end{equation}
where the integral domain is real axis in $\Omega_{+}$ given in
eq.(\ref{36}). We obtain
\begin{equation}\label{73}
    {\rm ln}a(\omega_3(\lambda))=I_0-iI_1\frac{\lambda}{2\rho}+\cdots
\end{equation}
where $I_0=0$ and
\begin{equation}\label{74}
    I_1=\frac{2}{\pi}\int_{\Omega_{+}} d \omega_3(\lambda') {{\rm
ln}|a(\omega_3(\lambda'))|^2}=\frac{2}{\pi}\int_{\Omega_{+}} d
\lambda' \frac{d\omega_3}{d\lambda'}{{\rm
ln}|a(\omega_3(\lambda'))|^2}
\end{equation}
It approaches to eq.(\ref{55}) in the domain of $|\lambda|\approx
0$.

\section{Discrete spectrum}

Eq.(\ref{63}) should include the discrete part
\begin{equation}\label{75}
    a_d(k)=\prod_j\frac{k-k_j}{k-\bar{k}_j}
\end{equation}
where $k_j$ in the complex $k-$plane. It can be transformed into
$\lambda-$plane, that is
\begin{equation}\label{76}
    a_d(\lambda)=\prod_j\frac{\lambda-\lambda_j}{\lambda-\bar{\lambda}_j}\frac{\bar{\lambda}_j}{\lambda_j}
\end{equation}
In general, we write
\begin{equation}\label{77}
    a_d(\omega_3(\lambda))=\prod_j\frac{\omega_3(\lambda)-\omega_3(\lambda_j)}
{\omega_3(\lambda)-\overline{\omega_3(\lambda_j)}}
\end{equation}
In the limit of $|\lambda|\approx0$, it coincides with eq.(\ref{76}) since
$\omega_3(\lambda)\to \rho \lambda^{-1}$ in the limit of $|\lambda|\to0$.

Extending analytically into $\Omega_{+}$, the left hand side of
eq.(\ref{39}) is
\begin{eqnarray}
    \{{\rm ln}a(\omega_3(\lambda)),\,b(\lambda_k)\}=\cdots+\sum_j
\{\big({\rm ln}(\omega_3(\lambda)-\omega_3(\lambda_j))-{\rm
ln}(\omega_3(\lambda)-\overline{\omega_3(\lambda_j)})\big),\,b(\lambda_k)\}\nonumber\\
=\cdots-\sum_j\frac{1}{\omega_3(\lambda)-\omega_3(\lambda_j)}\{\omega_3(\lambda_j),\,b(\lambda_k)\}
+\sum_j\frac{1}{\omega_3(\lambda)-\overline{\omega_3(\lambda_j)}}\{\overline{\omega_3(\lambda_j)},\,b(\lambda_k)\}
\label{78}
\end{eqnarray}
where $\lambda_k,\lambda_j\in \Omega_{+}$ and  the right hand side
is
\begin{equation}\label{79}
    \frac{\omega_1(\lambda)\omega_2(\lambda_k)+\omega_2(\lambda_k)\omega_1(\lambda)}{2}
\frac{1}{\omega_3(\lambda)-\omega_3(\lambda_k)}b(\lambda_k)
\end{equation}
Eq.(\ref{79}) has a pole at $\omega_3(\lambda)=\omega_3(\lambda_k)$ so
that
\begin{equation}\label{80}
    \{\omega_3(\lambda_j),\,b(\lambda_k)\}
=-\frac{\omega_1(\lambda_j)\omega_2(\lambda_k)+\omega_2(\lambda_k)\omega_1(\lambda_j)}{2}
b(\lambda_k)\delta_{jk}
=\rho\frac{d\omega_3}{d\lambda}\Big|_{\lambda_j}
b(\lambda_k)\delta_{jk}
\end{equation}
as seen in eq.(\ref{43}). Then eq.(\ref{80}) gives
\begin{equation}\label{81}
    \{\lambda_j,b(\lambda_k)\}=b(\lambda_k)\delta_{jk}
\end{equation}

\section{Action-angle variables in discrete spectrum}

We introduce the action-angle variables in discrete spectrum
\begin{equation}\label{82}
    P_j=F(\omega_3(\lambda_j)),~~Q_j={\rm ln}b(\omega_3(\lambda_j))
\end{equation}
Then we find
\begin{equation}\label{83}
    \{P_j,b(\lambda_k)\}=\frac{dF}{d\omega_3(\lambda_j)}\{\omega_3(\lambda_j),\,b(\lambda_k)\}
=\frac{dF}{d\omega_3(\lambda_j)}\frac{d\omega_3}{d\lambda_j}b(\lambda_k)\delta_{jk}
\end{equation}
namely,
\begin{equation}\label{84}
    \frac{dF}{d\omega_3(\lambda_j)}\frac{d\omega_3}{d\lambda_j}=1
\end{equation}
and then
\begin{equation}\label{85}
    P(\omega_3(\lambda_j))=\lambda_j
\end{equation}

From eq.(\ref{77}) the discrete part of Hamiltonian is
\begin{equation}\label{86}
    H_d=i4\rho\sum_n\left(-\omega_3(\lambda_n)+\overline{\omega_3({\lambda}_n)}\right)
\end{equation}
Here $\lambda_n$ lies in $\Omega_{+}$, and the factor 4 stands for
four zeros of $a(\lambda)$ in $\Omega$ for a single soliton
case. We thus obtain
\begin{equation}\label{87}
    \{H_d,\,b(\lambda_k)\}=i4\rho{\frac{d\omega_3}{d\lambda_j}}\{\lambda_j,\,b(\lambda_k)\}
=-i4{\omega_1\omega_2}b(\lambda_k)
\end{equation}
which is just canonical equation for $b(\lambda_k)$.

\section{Concluding remarks}

The Hamiltonian theory to the L-L equation for a spin chain with
full anisotropy was examined by E.K.Sklyanin$^2$. As mentioned,
the conservative quantities derived by him are not compatible with
the usual conservative quantities, such as the Hamiltonian, which
affirmatively concludes that his procedure is unreasonable.
Moreover, some other results may not be beyond doubt. For example,
his eq.(2.15) written in the present notation is
\begin{equation}\label{88}
    \{a(\lambda),b(\lambda')\}=-\omega_3(\lambda-\lambda'+i0)a(\lambda)b(\lambda')
\end{equation}
which is questionable, as comparing with eqs.(\ref{39}) or (\ref{44}).

Mikhailov and Rodin tried to solve the Landau-Lifschitz equation
for a spin chain with full anisotropy based upon the compatibility
pair given by Sklyanin. But explicit solutions were not given. In the
isotropic spin case, the Jost solutions do not approach
to the free Jost solutions as spectral parameter $|k|\to\infty$.
To construct the equations of the inverse scattering transform by Cauchy
contour integral, Takhtajan introduced a redundant factor $k^{-1}$
to ensure the integral having vanishing contribution of the
integral along the big circle in complex $k-$plane when the radius
reaches infinity. In the case of spin chain with full
anisotropy, the behaviors of the Jost solutions do not approach to
the free Jost solutions when $|\lambda|\approx0$. Rodin and
Mikhailov are unable to overcome this difficulty\cite{3,4}. But in order to solve the
equation, it is necessary to propose a way to do so.

\begin{acknowledgments}
The National Natural Science Foundation of China under Grant No.
10375041 supports the project.
\end{acknowledgments}

\appendix

\section{}

To calculate $f_{\alpha}$ in (26), the terms involving $S_3$,
$S_1$, $S_2$ in eq.(26) are
\begin{eqnarray}
    -(\omega_1\omega'_2\sigma_1\otimes\sigma_2-\omega_2\omega'_1\sigma_2\otimes\sigma_1)
&=&(f_3-f_0)i(\omega_3-\omega'_3)(I\otimes'\sigma_3-\sigma_3\otimes'I)\nonumber\\
&~&+(f_1-f_2)i(\omega_3+\omega'_3)(i\sigma_1\otimes'\sigma_2+i\sigma_2\otimes'\sigma_1)\label{a1}
\end{eqnarray}
\begin{eqnarray}
    -(\omega_2\omega'_3\sigma_2\otimes\sigma_3-\omega_3\omega'_2\sigma_3\otimes\sigma_2)
&=&(f_1-f_0)i(\omega_1-\omega_1')(I\otimes'\sigma_1-\sigma_1\otimes'I)\nonumber\\
&~&+(f_2-f_3)i(\omega_1+\omega_1')(i\sigma_2\otimes'\sigma_3+i\sigma_3\otimes'\sigma_2)\label{a2}
\end{eqnarray}
\begin{eqnarray}
-(\omega_3\omega'_1\sigma_3\otimes\sigma_1-\omega_1\omega'_3\sigma_1\otimes\sigma_3)
&=&(f_2-f_0)i(\omega_2-\omega_2')(I\otimes'\sigma_2-\sigma_2\otimes'I)\nonumber\\
&~&+(f_3-f_1)i(\omega_2+\omega_2')(i\sigma_3\otimes'\sigma_1+i\sigma_1\otimes'\sigma_3)\label{a3}
\end{eqnarray}
among which one equation turns to one another by simple circular permutation of $1,2,3$.
Noticing the direct products in two sides are different, and writing
them in $4\times4$ matrix form, we obtain
\begin{equation}\label{a4}
    \omega_1\omega'_2-\omega_2\omega'_1
=2(f_1-f_2)(\omega_3+\omega'_3),~~
\omega_1\omega'_2+\omega_2\omega'_1
=2(f_3-f_0)(\omega_3-\omega_3')
\end{equation}
\begin{equation}\label{a5}
    \omega_2\omega'_3-\omega_3\omega'_2
=2(f_2-f_3)(\omega_1+\omega'_1),~~
\omega_2\omega'_3+\omega_3\omega'_2
=2(f_1-f_0)(\omega_1-\omega_1')
\end{equation}
\begin{equation}\label{a6}
    \omega_3\omega'_1-\omega_1\omega'_3
=2(f_3-f_1)(\omega_2+\omega'_2),~~
\omega_3\omega'_1+\omega_1\omega'_3
=2(f_2-f_0)(\omega_2-\omega_2')
\end{equation}

\section{}

Denoting
\begin{equation}\label{a7}
    A_{\alpha}(L)=\Psi^{-1}(L,\lambda)\sigma_{\alpha}\Psi(L,\lambda'),~~
A_{\alpha}(-L)=\Psi^{-1}(-L,\lambda)\sigma_{\alpha}\Psi(-L,\lambda')
\end{equation}
\begin{equation}\label{a8}
    C_{\alpha}(L)=\Phi^{-1}(L,\lambda')\sigma_{\alpha}\Phi(L,\lambda),~~
C_{\alpha}(-L)=\Phi^{-1}(-L,\lambda')\sigma_{\alpha}\Phi(-L,\lambda)
\end{equation}
We can see
\begin{eqnarray}
&&\Delta_0\equiv A_0(L)\otimes'
C_0(L)-A_0(-L)\otimes'C_0(-L)=\Delta(b)+\Delta(b0)\label{a9-1}\\
&&\Delta_3\equiv A_3(L)\otimes'
C_3(L)-A_3(-L)\otimes'C_3(-L)=\Delta(b)-\Delta(b0)\label{a9-2}\\
&&\Delta_1\equiv A_1(L)\otimes'
C_1(L)-A_1(-L)\otimes'C_1(-L)=-\Delta(b)+\Delta(b1)\label{a9-3}\\
&&\Delta_2\equiv A_2(L)\otimes'
C_2(L)-A_2(-L)\otimes'C_2(-L)=-\Delta(b)-\Delta(b1)\label{a9-4}
\end{eqnarray}
where
\begin{equation}\label{a10}
    \Delta(b)\equiv\left(\!\!\begin{array}{cccc} 0&-a\tilde{b}'&
-\tilde{a}'\tilde{b}&0\\ [2pt] -b'a&0& 0&\tilde{b}a'\\ [2pt]
-b\tilde{a}'&0&0& \tilde{b}'\tilde{a}\\ [2pt]
0&a'b&\tilde{a}b'& 0
\end{array}\!\!\right)
\end{equation}
\begin{equation}\label{a11}
    \Delta(b0)\equiv\left(\!\!\begin{array}{cccc}
C_{11,11}&\tilde{b}a'e^{i2(\omega_3-\omega_3')
L}&\tilde{b}'\tilde{a}e^{-i2(\omega_3'-\omega_3) L}&0\\ [6pt]
a'be^{-i2(\omega_3'-\omega_3) L}&0&C_{12,21}&
-a\tilde{b}'e^{-i2(\omega_3-\omega_3') L}\\ [6pt]
\tilde{a}b'e^{i2(\omega_3-\omega_3') L}&C_{21,12}&
0&-\tilde{a}'\tilde{b}e^{i2(\omega_3'-\omega_3) L}\\ [6pt]
0&-b'ae^{i2(\omega_3'-\omega_3)
L}&-b\tilde{a}'e^{-i2(\omega_3-\omega_3') L}& C_{22,22}
\end{array}\!\!\right)
\end{equation}
\begin{eqnarray}
&&C_{11,11}=\tilde{b}'be^{-i2(\omega_3'-\omega_3)L}-\tilde{b}b'e^{i2(\omega_3-\omega_3')L}\nonumber\\
&&C_{12,21}=a'\tilde{a}e^{-i2(\omega_3'-\omega_3) L}-a\tilde{a}'e^{-i2(\omega_3-\omega_3') L}\nonumber\\
&&C_{21,12}=\tilde{a}'ae^{i2(\omega_3'-\omega_3) L}-\tilde{a}a'e^{i2(\omega_3-\omega_3') L}\nonumber\\
&&C_{22,22}=b'\tilde{b}e^{i2(\omega_3'-\omega_3)L}-b\tilde{b}'e^{-i2(\omega_3-\omega_3')L}\label{a12}
\end{eqnarray}
and
\begin{equation}\label{a13}
    \Delta(b1)\equiv \left(\begin{array}{cccc}
0 &\tilde{a}'be^{i2(\omega_3+\omega_3') L} &ab'e^{-i2(\omega_3+\omega_3') L} &C_{11,22}\\
[6pt] \tilde{b}\tilde{a}'e^{i2(\omega_3+\omega_3') L} &C_{12,12}
&0& -b'\tilde{a}e^{i2(\omega_3+\omega_3') L}\\ [6pt]
\tilde{b}'ae^{-i2(\omega_3+\omega_3') L} &0 & C_{21,21}
&-ba'e^{-i2(\omega_3+\omega_3') L}\\ [6pt] C_{22,11}
&-\tilde{a}\tilde{b}'e^{i2(\omega_3+\omega_3') L}
&-a'\tilde{b}e^{-i2(\omega_3+\omega_3')L} & 0
\end{array}\right)
\end{equation}
\begin{eqnarray}
&&C_{11,22}=\tilde{a}'\tilde{a}e^{i2(\omega_3+\omega_3') L}-aa'e^{-i2(\omega_3+\omega_3') L}\nonumber\\
&&C_{12,12}=-b'be^{i2(\omega_3+\omega_3') L}+\tilde{b}\tilde{b}'e^{i2(\omega_3+\omega_3') L}\nonumber\\
&&C_{21,21}=-\tilde{b}'\tilde{b}e^{-i2(\omega_3+\omega_3')L}+bb'e^{-i2(\omega_3+\omega_3') L}\nonumber\\
&&C_{22,11}=a'ae^{-i2(\omega_3+\omega_3')L}-\tilde{a}\tilde{a}'e^{i2(\omega_3+\omega_3') L}\label{a14}
\end{eqnarray}

Here we show the procedure for obtaining eq.(\ref{a9-1}) as an example.
Firstly, there is
\begin{eqnarray}
    A_0(L)\otimes'
C_0(L)&=&\left(\begin{array}{cc}e^{i(\omega_3-\omega_3') L}&0\\
0&e^{-i(\omega_3-\omega_3') L}\end{array}\right)\label{a15}\\
&~&\otimes'\left(\begin{array}{cc}\tilde{a}'ae^{i(\omega_3'-\omega_3)L}+\tilde{b}'be^{-i(\omega_3'-\omega_3)
L}&
-\tilde{a}'\tilde{b}e^{i(\omega_3'-\omega_3)L}+\tilde{b}'\tilde{a}e^{-i(\omega_3'-\omega_3)
L}\\
-b'ae^{i(\omega_3'-\omega_3)L}+a'be^{-i(\omega_3'-\omega_3) L}&
b'\tilde{b}e^{i(\omega_3'-\omega_3)L}+a'\tilde{a}e^{-i(\omega_3'-\omega_3)
L}\end{array}\right) \nonumber
\end{eqnarray}
According to the definition of $\otimes'$ in eq.(\ref{24}), the terms
involving vanishing exponent $e^0$ and the terms involving
non-vanishing exponent $e^{\pm i2(\omega_3-\omega_3')L}$  are collected
separately,
\begin{equation}\label{a16}
    \left(\!\!\begin{array}{cccc} \tilde{a}'a&0&
-\tilde{a}'\tilde{b}
&0\\
-b'a&0&
b'\tilde{b}&0\\
0&\tilde{b}'b&0&
\tilde{b}'\tilde{a}\\
0&a'b&0& a'\tilde{a}
\end{array}\!\!\right)+
\left(\!\!\begin{array}{cccc}
\tilde{b}'be^{-i2(\omega_3'-\omega_3) L}&0&
\tilde{b}'\tilde{a}e^{-i2(\omega_3'-\omega_3) L}
&0\\
a'be^{-i2(\omega_3'-\omega_3) L}&0&
a'\tilde{a}e^{-i2(\omega_3'-\omega_3) L}&0\\
0&\tilde{a}'ae^{i2(\omega_3'-\omega_3) L}&0&
-\tilde{a}'\tilde{b}e^{i2(\omega_3'-\omega_3) L}\\
0&-b'ae^{i2(\omega_3'-\omega_3) L}&0&
b'\tilde{b}e^{i2(\omega_3'-\omega_3) L}
\end{array}\!\!\right)
\end{equation}
Secondly,
\begin{eqnarray}
    A_0(-L)\otimes'C_0(-L)&=&
\left(\begin{array}{cc}a\tilde{a}'e^{-i(\omega_3-\omega_3')
L}+\tilde{b}b'e^{i(\omega_3-\omega_3') L}&
a\tilde{b}'e^{-i(\omega_3-\omega_3')
L}-\tilde{b}a'e^{i(\omega_3-\omega_3') L}\\ [6pt]
ba'e^{-i(\omega_3-\omega_3')
L}-\tilde{a}b'e^{i(\omega_3-\omega_3') L}&
b\tilde{b}'e^{-i(\omega_3-\omega_3') L}+\tilde{a}a'e^{i(\omega_3-\omega_3') L}\end{array}\right)\nonumber\\
&~&\otimes'\left(\begin{array}{cc}e^{-i(\omega_3'-\omega_3) L}&0\\
0&e^{i(\omega_3'-\omega_3) L}\end{array}\right)\label{a17}
\end{eqnarray}
is equal to
\begin{equation}
    \left(\begin{array}{cccc} a\tilde{a}'&
a\tilde{b}'&0&0\\
0&0&\tilde{b}b'
&-\tilde{b}a'\\
b\tilde{a}'&
b\tilde{b}'&0&0\\
0&0&-\tilde{a}b'& \tilde{a}a'
\end{array}\right)+\nonumber\\
\left(\begin{array}{cccc} \tilde{b}b'e^{i2(\omega_3-\omega_3')
L}&
-\tilde{b}a'e^{i2(\omega_3-\omega_3') L}&0&0\\
0&0&a\tilde{a}'e^{-i2(\omega_3-\omega_3') L}
&a\tilde{b}'e^{-i2(\omega_3-\omega_3') L}\\
-\tilde{a}b'e^{i2(\omega_3-\omega_3') L}&
\tilde{a}a'e^{i2(\omega_3-\omega_3') L}&0&0\\
0&0&b\tilde{a}'e^{-i2(\omega_3-\omega_3') L}&
b\tilde{b}'e^{-i2(\omega_3-\omega_3') L}
\end{array}\right)\label{a18}
\end{equation}
Finally, combining (\ref{a16}) and (\ref{a18}), we have shown eq.(\ref{a9-1}).


\begin{thebibliography}{100}
\bibitem{1} L.D.Landau and E.M.Lifschitz,(1935). In: Collected Papers of
Landau L.D. ed. D.ter Haar, p.101. New York, Pergamon, Gordon and
Breach (1965).
\bibitem{2}E.K.Sklyanin, Preprint LOMI K-9-79, Liningrad (1979).
\bibitem{3}A.V.Mikhailov, Physica 3D,73(1981); Phys.Lett. 92A,51(1982).
\bibitem{4}Yu.L.Rodin, Lett.Math.Phys. 7,3(1983).
\bibitem{5}L.D.Faddeev and L.A.Takhtajan, Hamiltonian Method in the
Theory of Solitons(Springer, Berlin,1987).
\bibitem{6}J.C.He, L.N.Shi,H.Chen and N.N.Huang, J.Phys.A37,6311(2004).
\bibitem{7}V.E.Zakharov and L.A.Takhtajan, Theor.Math.Phys.38,17(1979).
\bibitem{8}N.N.Huang, H.Cai, F.M.Liu, L.N.Shi, Chin.Phys.Lett.21,1699(2004)
\end{thebibliography}
\end{document}